# Pre-processing methods for nodule detection in lung CT

P. Delogu[a,*], S.C. Cheran[b], I. De Mitri[c], G. De Nunzio[d], M.E. Fantacci[a], F. Fauci[e], G. Gargano[f], E. Lopez Torres[g], R. Massafra[f], P. Oliva[h], A. Preite Martinez[a], G. Raso[e], A. Retico[a], S. Stumbo[h], A. Tata[a]

[a]*Dipartimento di Fisica dell'Università and Sezione I.N.F.N., Pisa, Italy*
[b]*Dipartimento di Informatica dell'Università and Sezione I.N.F.N., Torino, Italy*
[c]*Dipartimento di Fisica dell'Università and Sezione I.N.F.N., Lecce, Italy*
[d]*Dipartimento di Scienza dei Materiali dell'Università and Sezione I.N.F.N., Lecce, Italy*
[e]*Dipartimento di Fisica e Tecnologie Relative dell'Università di Palermo and Sezione I.N.F.N. di Catania, Italy*
[f]*Dipartimento di Fisica dell'Università and Sezione I.N.F.N., Bari, Italy*
[g]*CAEDEN, Havana, Cuba*
[h]*Struttura Dipartimentale di Matematica e Fisica dell'Università di Sassari and Sezione I.N.F.N. di Cagliari, Italy*

**Abstract.** The use of automatic systems in the analysis of medical images has proven to be very useful to radiologists, especially in the framework of screening programs, in which radiologists make their first diagnosis on the basis of images only, most of those corresponding to healthy patients, and have to distinguish pathological findings from non-pathological ones at an early stage. In particular, we are developing preprocessing methods to be applied for pulmonary nodule Computer Aided Detection in low-dose lung Multi Slice CT (Computed Tomography) images.

## 1. Introduction

It has been proven that early detection and resection of lung cancer can improve the prognosis significantly [1]: the overall 5-year survival rate of 14% increases to 49% if the lesion is localized and decreases to 2% if it has metastasized and for Stage I cancer is 67%. The problem is that early curable lung cancers usually produce no symptoms and are often missed in mass screening programs in which chest radiography, which has been used for detecting lung cancer for a long time, is employed [2]: the results of screening trials with chest radiography and cytological examination of sputum were interpreted as failing to demonstrate a reduction in mortality [3]. Computer Tomography (CT) has been shown to be the best imaging modality for detection of pulmonary nodules [5], particularly since the introduction of spiral (helical) technology. The CT sensitivity increases with a reduction of the slice thickness due to a decrease of partial-volume effect, and overlapping images reconstruction improves the detection of small nodules located at the boundary of two contiguous non-overlapping images. Typical chest CT protocols are, however, associated with relatively high radiation exposure to the patient [4], which causes concern about induction of malignant disease, particularly in a screening setting. Many studies [4] have suggested that dose reduction in CT does not decrease its sensitivity for small pulmonary nodules. The efficacy of low dose helical CT protocols [3] has renewed the interest in (and also the demand for) lung cancer screening, although the effect of any lung cancer screening program, including screening with low-dose helical CT, remains an open issue in the medical community. Moreover, the number of images that need to be interpreted in CT screening is high [1], particularly when multi-detector helical CT and thin collimation are used, and most of them correspond to non-pathological patients. Since missed cancers are not uncommon also in CT interpretation [1], double reading would be useful, but it doubles the demand of radiologists' time. So researchers have recently begun to explore CAD methods in this area [1,6].

---
[*] Corresponding author. *E-mail address*: delogu@df.unipi.it

The work presented in this paper as been carried out in the framework of the INFN (Istituto Nazionale di Fisica Nucleare)-funded MAGIC-5 (Medical Applications in a GRID Infrastructure Connection) experiment, whose overall goal is the realization of software for medical applications as Computer Aided Detection (CAD) Systems and distributed databases to be implemented in a GRID Virtual Organization. We present an overview about the methods that we have developed for pulmonary nodule Computer Aided Detection in CT images and the preliminary results obtained.

## 2. Methods

Generally speaking, there are two possible philosophies for designing a CAD system: the first is based on the determination (by means of thresholding and segmentation algorithms) of the parenchymal volume and the search of the nodules only inside that volume; the second is based on the search of the nodules in the overall CT-scanned volume and in the determination of the position of the nodule with respect to the parenchyma. We developed and tested some algorithms and evaluated both the advantages and the disadvantages of the two possible approaches in view of the final implementation in a complete CAD system. The requirements for a good performing CAD following the first strategy are: to define exactly the volumes (because a not-exactly defined parenchymal volume will lead a worse sensitivity of the CAD); to have a good nodule hunter. On the other hand the disadvantage in the second strategy is the possibility of a fall in the specificity performances. Another aspect to be taken into account differently in the two approaches is related to the position of the nodules. In the first strategy it is possible to define two classes of nodules: the first one includes the "internal" nodules, i.e. those located inside the lung volume, which are quite spherical; the other class includes the nodules located very close to the pleura, witch are characterized by a more irregular shape. In the second strategy, the nodule-detection algorithm has to be sufficiently robust to allow the detection of this last kind of nodules. Moreover, an important and difficult problem for nodule detection is the selection of the nodule candidates. This is a very difficult task since some nodules [6] may be characterized by very low CT values and low contrast (especially those with ground-glass opacity), may have CT values similar to those of blood vessels and airway walls, or may be connected to blood vessels and airway walls. In our work we have used both the approaches, using a same nodule detection algorithm.

In the first approach, the first step is based on a combination of image-processing techniques, such as threshold-based segmentation, morphological operators, border detection, border thinning, border reconstruction, and region filling. The second step detects nodule candidates based on a dot-enhancement filter. We first transform the slices into the PGM (Portable Gray Map) format using DCMTK, which is a collection of C/C++ libraries and applications that implement the DICOM standard. In order to extract the pulmonary parenchyma, we perform the following actions. First we remove the background (i.e., the pixels with the same grey level as the lungs but located outside the chest) from the image; then we generate a binary image by means of a thresholding technique that uses either a static or a dynamic threshold depending on the lung zone the slice belongs to. Afterwards, we apply the morphological opening and closing operators so as to improve the image and border definition; this allows us, for instance, to enhance the separation between distinct regions and to fill the gaps in the borders. Then we detect the image borders through a tracking algorithm that uses the Sobel operator, and reduce the border size to one pixel using the Wang and Zhang thinning algorithm [7]. After identifying the border pixel chains that represent the two pulmonary lobes based on both the chain size and the chain location within the slices, we reconstruct the pulmonary lobes borders so as to reinsert the nodules adjacent to the pleura previously suppressed by the thresholding operator. Finally, we apply a region filling operator to the pulmonary lobes chains in order to reintroduce the original values of grey levels inside the lungs. The resulting images, which contain only the lung regions, are input to the nodule detection filter.

In the nodule candidate detection step, common to both the approaches, a dot-enhancement filter is applied to the 3D matrix of voxel data. This 3D filter attempts to determine local geometrical characteristics for each voxel, computing the eigenvalues of the Hessian matrix and evaluating a "likelihood" function that was purposely built to discriminate between local morphology of linear, planar and spherical objects, modeled as having 3D Gaussian sections (Q. Li, S. Sone and K. Doi [6]). By applying this 3D filter to artificial images, we have verified the efficiency in detecting the Gaussian-like regions even in the cases were they are superimposed to non-Gaussian ones (see Fig. 1). This is a very important property of the 3D filter since it can allow for the detection of nodules located near to the pleura, blood vessels or airway walls.

A simple peak-detection algorithm (i.e. a local maximum detector) is then applied to the filter output, in order to detect the filtered signal peaks.

Finally, in the second approach, the position of the nodule candidate with respect to the parenchyma is evaluated by rejecting the candidates on a background with CT numbers far away from typical values that characterize the lung. The Fig. 2 shows a typical output of the 3D filter.

We tested the algorithms on a dataset of standard and low dose CT scans. In particular, we applied this software system to sets of lung multi-slice CT scans acquired at high (standard setting: 120 kV, 120 mA) and low (screening setting: 120 kV, 20 mA) dose with reconstructed thickness of 5 mm and 1 mm, respectively. The database consists of 4 high dose scans and 4 low dose ones of the same patients, with a total of 8 nodules annotated by an experienced radiologist. The linear size of these nodules range from 0.5 to 5 mm and 5 of them are located inside the lung volume, while the remaining 3 are located very close to the pleura. Each scan is a sequence of about 300 slices stored in DICOM (Digital Imaging and COmmunications in Medicine) format.

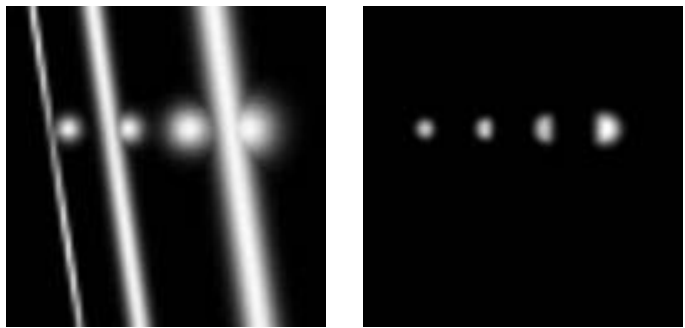

Fig. 1. Artificial image (left) and output of the dot enhancement filter (right).

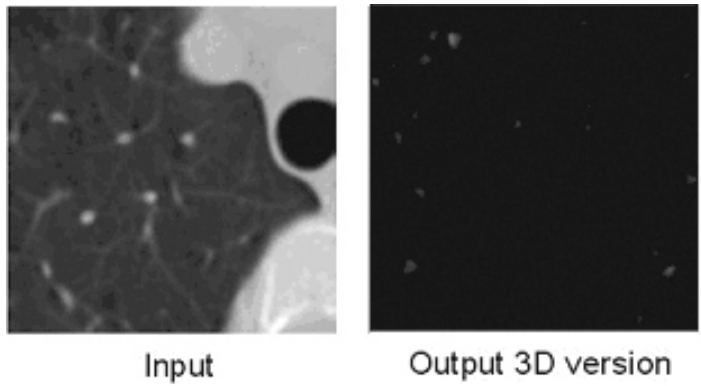

Fig. 2. Original image (left), output of the 3D (right) dot-enhancement filter in a non-pathological case.

## 3. Results and conclusions

Our preliminary results show that the 3D filter applied following the second approach is characterized by a high sensitivity. In particular, both the 5 internal than the 3 nodules close to the pleura are detected. This result is obtained in both the standard dose images and the low dose ones, proving that the system we developed is acceptable as a pre-analysis stage of nodule detection even for low dose images. We are currently working on the reduction of the amount of the false positives findings. In order to achieve this goal, we are developing a 3D algorithm based on the wavelet transforms.

**Acknowledgments**

We would like to acknowledge Dr. L. Battolla, Dr. F. Falaschi and Dr. C. Spinelli of the Unità Operativa Radiodiagnostica 2 of the Azienda Universitaria Pisana and Prof. D. Caramella and Dr. T. Tarantino of the Diagnostic and Interventional Radiology Division of the Dipartimento di Oncologia, Trapianti e Nuove Tecnologie in Medicina of the Pisa University who have made available the full annotated CT scans.

**References**

[1] M.N. Gurcan, B. Sahinder, N. Petrick et al., Lung nodule detection on thoracic computed tomography images: preliminary evaluation of a computer-aided diagnosis system, Med. Phys. 29 (11) (2002), pp. 2552-2558.
[2] S. Itoh, M. Ikeda, S. Arahata et al., Lung cancer screening: minimum tube current required for helical CT, Radiology 215 (2000), pp. 175-183.
[3] S.G. Armato, F. Li, M.L. Giger, H. Macmahon, S. Sone, K. Doi, Lung cancer: performance of automated lung nodule detection applied to cancers missed in a CT screening program, Radiology 225 (2002), pp. 685-692.
[4] S. Diederich, D. Wormanns, M. Semik et al., Screening for early lung cancer with low-dose spiral CT: prevalence in 817 asymptomatic smokers, Radiology 222 (2002), pp. 773-781.
[5] S. Diederich, M.G. Lentschig, T.R. Overbeck, D. Wonnanns, W. Hemdel, Detection of pulmonary nodule at spiral CT: comparison of maximum intensity projection sliding slabs and single-image reporting, Eur. Radiol. 11 (2001) pp. 1345-1350.
[6] Q. Li, S. Sone, K. Doi, Selective enhancement filters for nodules, vessels, and airways walls in two- and three- dimensional CT scans, Med. Phys. 30 (8) (2003) pp. 2040-2051.
[7] P.S.P. Wang, Y.Y. Zhang, A fast and flexible thinning algorithm, IEEE Transactions on Computers 38 (5) (1989), pp. 741-745.